\documentclass{article}
\usepackage{amssymb}
\usepackage{graphicx}
\usepackage{amsmath}

\input{tcilatex}

\begin{document}

\title{Kallen-Lehman approach to 3D Ising model}
\author{F. Canfora \\
Centro de Estudios Cientificos (\textit{CECS), Casilla 1469 Valdivia, Chile.}%
\\
\textit{Istituto Nazionale di Fisica Nucleare, GC di Salerno, Italy.}\\
}
\maketitle

\begin{abstract}
A "Kallen-Lehman" approach to Ising model, inspired by quantum field theory 
\textit{a la} Regge, is proposed. The analogy with the Kallen-Lehman
representation leads to a formula for the free-energy of the 3D model with
few free parameters which could be matched with the numerical data. The
possible application of this scheme to the spin glass case is shortly
discussed.

Keyword: Regge theory, spin glasses, Ising model.

PACS: 12.40.Nn,11.55.Jy, 05.20.-y, 05.70.Fh.
\end{abstract}

\section*{Introduction}

The three dimensional Ising model (henceforth 3DI) is one of the main open
problems in field theory and statistical mechanics. A large number of
interesting statistical systems near the transition point are described by
3DI and the theoretical methods suitable to deal with such a problem
manifest deep connections in various areas of physics ranging from quantum
information theory to string theory (to provide with even a partial list of
references is really a completely hopeless task, a detailed review-which, by
the way, does not cover all the approaches and results on the subject-is 
\cite{PV02}). Besides its intrinsic interest is statistical physics, since
the formulation of the Svetitsky-Yaffe conjecture \cite{SV82}, it has been
widely recognized its role in describing the deconfinement transition in
QCD. For this reason, the 3D Ising model is worth to be further investigated.

Here, a "phenomenological"\footnote{%
The meaning of the term "phenomenological" in this context will be more
clear later on.} approach is proposed which allows to formally connect the
3DI model to the Ising model in one dimension less. The method is inspired
by quantum field theory \textit{a la} Regge and by the Kallen-Lehman
representation. Recently (see \cite{CPV06}) this approach has been able to
shed some light on the physical origin of non extensivity.

The paper is organized as follows: in the second section the "Kallen-Lehman
interpretation" of the 2D Ising model will be discussed. In the third
section the 3D model will be described. In the fourth section the high
temperature behavior will be analyzed and it will be shortly pointed out how
to apply the formalism to the spin glass case. Eventually, some conclusion
will be drawn.

\section{Ising Kinetic term(s)}

In this section the main idea of the method are introduced.

The partition of functions of Ising models in various dimensions all belong
to the following class of partition functions%
\begin{eqnarray}
Z(J_{ab}) &=&\dsum\limits_{\sigma _{a}}\exp \left[ \sum_{ab}J_{ab}\sigma
^{a}\sigma ^{b}\right]  \label{part1} \\
\sigma ^{a} &=&\pm 1\quad \forall a,\quad a,b=1,..,N  \label{part2}
\end{eqnarray}%
where the sum is over all the spin configurations, $N$ is the total number
of spins in the system and the "kinetic matrix" $J_{ab}$ identifies the
models. A useful way to write the Ising kinetic matrix is as follows. Let us
begin with the 3DI model on a finite cube of $n\times n\times n$ size (where 
$n\gg 1$): one can enumerate the spins starting from the leftmost spin on
the lowermost face of the cube and continuing from left to right in any row,
from the lowest row to the highest row in any face, and from the lowest face
to the highest face in the cube. In such a way, the nearest neighbors of the
spin number $k$ are the spins with numbers $k\pm 1$, $k\pm n$ and $k\pm
n^{2} $. So that the kinetic matrix of the 3DI model can be written as
follows:%
\begin{eqnarray}
J_{ab}^{(3)} &=&\beta \left[ \delta (\left\vert a-b\right\vert -1)+\delta
(\left\vert a-b\right\vert -n)+\delta (\left\vert a-b\right\vert -n^{2})%
\right] ,  \label{k3d} \\
\beta &=&\varepsilon /K_{B}T,\quad \varepsilon >0  \label{tempe}
\end{eqnarray}%
$K_{B}$ being the Boltzmann constant and $\varepsilon >0$ in order to have a
ferromagnetic interaction. From this construction it is also clear that the
kinetic matrices of the 1DI \ and 2DI models are:%
\begin{eqnarray}
J_{ab}^{(1)} &=&\beta \delta (\left\vert a-b\right\vert -1),  \label{k1d} \\
J_{ab}^{(2)} &=&\beta \left[ \delta (\left\vert a-b\right\vert -1)+\delta
(\left\vert a-b\right\vert -n)\right] .  \label{k2d}
\end{eqnarray}%
At least at a formal level, one can see the kinetic matrix of a
D-dimensional Ising model as a perturbation of the kinetic matrix of the
(D-1)-dimensional Ising model in which a further off-diagonal term has been
added. This point of view is useful since already the relation between the
2DI and the 1DI models fits with this scheme. To see this, it is worth to
write down the well known free energies of both models:%
\begin{eqnarray}
F_{1D}(\beta ) &=&\log 2\cosh \beta ,  \label{parto1} \\
F_{2D}(\beta ) &=&F_{1D}(2\beta )+\frac{1}{2\pi }\dint\limits_{0}^{\pi
}dx\log \left\{ \frac{1}{2}\left[ 1+\sqrt{1-k(\beta )^{2}\sin ^{2}x}\right]
\right\}  \label{parto2} \\
k(\beta ) &=&\frac{2}{\cosh 2\beta \coth 2\beta },\quad 0\leq k(\beta )\leq
1.  \label{parto3}
\end{eqnarray}%
The above interpretation of the 2DI model as a perturbation of the 1DI model
is particularly clear: the free energy $F_{2D}$ appears as a "dressing" 
\textit{a la} Kallen-Lehman (see, for example, \cite{We96}) of the partition
function $F_{1D}$. It is worth to recall here that the singularity in the
specific heat is determined by the region in which $k$ (which multiplies the 
$\sin $ function inside the square root) is near to $1$.

To be more specific, in a scalar interacting quantum field theory in which
the bare propagator $\Delta _{\phi }$ is%
\begin{equation*}
\Delta _{\phi }(m^{2})=\frac{-i}{p^{2}+m^{2}-i\varepsilon },
\end{equation*}%
the dressed propagator $\mathbf{\Delta }_{\phi }$ can be written as the
convolution of the bare propagator with a suitable energy function $\rho $:%
\begin{equation*}
\mathbf{\Delta }_{\phi }=\dint \rho _{\phi }(\mu ^{2})\Delta _{\phi }(\mu
^{2})d\mu ^{2}.
\end{equation*}%
Usually, the function $\rho _{\phi }$ has $\delta $ contributions (coming
from single particles states) plus other terms (coming from the continuous
spectrum)%
\begin{equation*}
\rho _{\phi }(\mu ^{2})\sim \dsum\limits_{i}\delta (\mu ^{2}-\alpha
m_{i}^{2})+\sigma _{\phi }(\mu ^{2}).
\end{equation*}

In very much the same way as it happens in field theory (in which the bare
propagator is dressed by the interactions which lead to the exact propagator
which can be expressed as a convolution of the bare propagator with a
suitable energy-density function), in this case $F_{2D}$ appears as a
convolution of $F_{1D}$ with a suitable density $\rho _{\beta }$:%
\begin{eqnarray}
F_{2D}(\beta ) &=&\int_{0}^{\infty }\rho _{\beta }(\mu )F_{1D}(\mu )d\mu ,
\label{dre1} \\
\rho _{\beta }(\mu ) &=&\delta (\mu -2\beta )+\sigma _{\beta }(\mu )
\label{dre2} \\
\sigma _{\beta }^{(\alpha )}(\mu ) &=&\frac{1}{2\pi }\dint\limits_{0}^{\pi
}d\phi J(\mu )\delta \left\{ \cosh \mu -\frac{1}{4}\left[ 1+\left( 1-k(\beta
)^{2}\sin ^{2}\phi \right) ^{\alpha }\right] \right\} ,  \label{dre3} \\
\alpha &=&\frac{1}{2}  \label{dre2D}
\end{eqnarray}%
where $J(\mu )$ is the Jacobian needed to have a well-defined $\delta $%
-function, the function $k(\beta )$ is defined in Eq. (\ref{parto3}). Thus,
the above arguments lead to think that a similar representation could also
hold in the case of the 3DI model in terms of the 2DI model.

It is worth to stress that the above dressing function $\sigma _{\beta
}^{(\alpha )}(\mu )$ can be thought as an infinite sum of contributions of
"moving poles": since%
\begin{equation*}
\delta (x)\sim \underset{\gamma \rightarrow 0}{\lim }P\left\{ \frac{\gamma }{%
x^{2}+\gamma ^{2}}\right\}
\end{equation*}%
(where $P\left\{ {}\right\} $ denotes the principal value) one can write%
\begin{equation*}
\sigma _{\beta }^{(\alpha )}(\mu )\sim \underset{\gamma \rightarrow 0}{\lim }%
\frac{1}{2\pi }P\dint\limits_{0}^{\pi }d\phi J(\mu )\left\{ \frac{\gamma }{%
\gamma ^{2}+\left\{ \cosh \mu -\frac{1}{4}\left[ 1+\left( 1-k(\beta
)^{2}\sin ^{2}\phi \right) ^{\alpha }\right] \right\} ^{2}}\right\}
\end{equation*}%
which looks like the sum of contributions of particles with $\phi $%
-dependent mass. The $\phi $-dependent "trajectories" of the (generically
complex) poles are determined implicitly by the following equations%
\begin{equation*}
\gamma ^{2}+\left\{ \cosh \mu -\frac{1}{4}\left[ 1+\left( 1-k(\beta
)^{2}\sin ^{2}\phi \right) ^{\alpha }\right] \right\} ^{2}=0
\end{equation*}%
(where, for the 2D Ising model, $\alpha =1/2$) and one has to sum (actually,
one has an integral over $\phi $) over the "moving poles". These "moving
poles" are indeed a characteristic features of Regge theory \cite{Re1} \cite%
{Re2}. The strength of the Regge approach to particles physics is its
generality \cite{Co71}\ \cite{Ve74}: it is based upon very general
assumption, it does not lie on perturbation theory and most of its results
(such as the Veneziano amplitude \cite{Ve1}) con be derived without a
precise knowledge of the microscopic interaction. It allows to describe in
terms of few parameters\footnote{%
Such parameters in the case of the old-fashoned S-matrix approach to the
strong interactions, were deduced by experimental data \cite{Co71}\ \cite%
{Ve74} but in principle could be computed in the underlying fundamental
theory. The strength of the Regge approach is that it says many things even
if one is not able to compute the above few parameters.} a huge number of
observations. The application of Regge theory to statistical mechanics has
been proposed for the first time in \cite{CPV06}.

\section{The "Kallen-Lehman" free energy}

In the previous section it has been suggested that the free energy of the
3DI model should be seen as a dressing of the free energy of the 2DI model.
At a first glance, there is no reason to expect that the dressing function
leading to $F_{3D}$ could bear any resemblance with the dressing function in
Eq. (\ref{dre2}) (which leads to $F_{2D}$). However, the lift from the
kinetic matrix of the 1DI model to the kinetic matrix of the 2DI model is
very similar to the lift from the kinetic matrix of the 2DI model to the
kinetic matrix of the 3DI model. Thus, the function which dresses the 2DI
model giving rise to the 3DI model could be quite similar to the function
which dresses the 1DI model giving rise to the 2DI model. One is lead to the
following ansatze for the free energy of the 3DI model:%
\begin{eqnarray}
F_{3D}^{(\nu ,\lambda )} &=&\int_{0}^{\infty }\rho _{\beta }(\mu )F_{2D}(\mu
)d\mu \Rightarrow  \label{parto31} \\
F_{3D}^{(\nu ,\lambda )}(\beta ) &=&F_{2D}(\beta )+\frac{\lambda }{\left(
2\pi \right) ^{2}}\dint\limits_{0}^{\pi }d\phi \dint\limits_{0}^{\pi
}d\varphi \cdot  \notag \\
&&\cdot \log \left\{ \frac{1}{2}\left[ 1+\sqrt{1-\left[ 2\frac{\sqrt{\Delta
(\phi )-1}}{\Delta (\phi )}\right] ^{\nu }\sin ^{2}\varphi }\right] \right\}
,  \label{parto32} \\
\Delta (\phi ) &=&\left( 1+\left( 1-k(\beta )^{2}\sin ^{2}\phi \right)
^{\alpha }\right) ^{2},\quad 1\leq \left( \Delta (\phi )\right) ^{2}\leq
4\Rightarrow  \notag \\
0 &\leq &2\frac{\sqrt{\Delta (\phi )-1}}{\Delta (\phi )}\leq 1,
\label{parto5}
\end{eqnarray}%
where $\rho _{J}(\mu )$ is defined in Eq. (\ref{dre2}), the function $%
k(\beta )$ is defined in Eq. (\ref{parto3}). In a sense, from the point of
view of Regge theory, the above dressing is the most "gentle" way to go
beyond the integrability of the 2D model: one introduces a more complex
analytic structure but in a "educated" recursive manner. The present field
theoretical point of view only suggests an ansatze and the correct scheme
should be to leave $\nu $ and $\lambda $ as a free parameters to be fitted
on numerical data. It will be shown that in order to describe the critical
behavior it is enough the parameter $\alpha $, the role of the other
parameters will be also discussed. It is worth to note here that in this
case also the factor which multiplies $\sin ^{2}\varphi $ inside the square
root fulfils the condition (\ref{parto5}) analogous to the condition (\ref%
{parto3}) which gives rise to the singularity in the 2D case (see \cite{O44}%
). The qualitative behavior of $F_{3D}^{(\nu ,\lambda ,\alpha )}(\beta )$
both at low and at high $\beta $ is similar to the two dimensional case.

\subsection{The specific heat}

In the present parametrization (which is the analogous of the classic
parametrization of the two dimensional model \cite{O44}), the specific heat
is proportional to the second derivative\footnote{%
The first derivative of the free energy has no problem since the possible
divergences of the of the integrand (which occur when $k\rightarrow 1$
and/or when $K_{\nu }\rightarrow 1$) are compensated by the fact that when $%
k\rightarrow 1$ one has that $k^{\prime }\rightarrow 0$ as well as when $%
K_{\nu }\rightarrow 1$ one has that $c(\Delta )\rightarrow 0$ (where $K_{\nu
}$ and $c(\Delta )$\ are defined below in the main text).} of $F_{3D}^{(\nu
,\lambda )}$ with respect to $\beta $:%
\begin{equation*}
\partial _{\beta }^{2}F_{3D}^{(\nu ,\lambda =1)}(\beta )\sim C_{(2)}+C_{(3)}
\end{equation*}%
where $C_{(2)}$ is the two dimensional contribution and $C_{(3)}$\ the three
dimensional one. A trivial but cumbersome computation gives%
\begin{eqnarray}
C_{(3)} &=&\dint\limits_{0}^{\pi }d\phi \dint\limits_{0}^{\pi }d\varphi
\left\{ \left[ \Delta ^{\prime }c\left( \Delta \right) \right] ^{2}\left[ 
\frac{I_{2}}{1-f_{(\nu )}}-\frac{I_{1}}{2\left( 1-f_{(\nu )}\right) ^{3/2}}%
\right] \right. +  \notag \\
&&+\left. I_{1}\frac{\left[ \Delta ^{\prime \prime }c(\Delta )+\left( \Delta
^{\prime }\right) ^{2}\frac{\partial c(\Delta )}{\partial \Delta }\right] }{%
\sqrt{1-f_{(\nu )}}}\right\} ,  \label{sh1} \\
f_{(\nu )} &=&\left( \sin ^{2}\varphi \right) K_{\nu },\quad K_{\nu }=\left[
2\frac{\sqrt{\Delta (\phi )-1}}{\Delta (\phi )}\right] ^{\nu }  \notag \\
\Delta ^{\prime } &=&\partial _{\beta }\Delta ,\quad \Delta ^{\prime \prime
}=\partial _{\beta }\Delta ^{\prime },\quad I_{1}=\frac{1}{N},\quad I_{2}=%
\frac{1}{N^{2}},  \notag \\
c(\Delta ) &=&\left( \sin ^{2}\varphi \right) \nu 2^{\nu -1}\frac{\left(
2-\Delta \right) \left( \sqrt{\Delta -1}\right) ^{\nu -2}}{\Delta ^{2+\nu }}%
,\quad \partial _{\beta }f_{(\nu )}=c(\Delta )\Delta ^{\prime }  \notag \\
\frac{\partial c(\Delta )}{\partial \Delta } &=&-\frac{\nu 2^{\nu -1}}{%
\Delta ^{2+\nu }}\left( \sin ^{2}\varphi \right) \left[ -\left( \frac{\nu -2%
}{2}\right) \left( 2-\Delta \right) \left( \Delta -1\right) ^{\frac{\nu -4}{2%
}}+\right.  \notag \\
&&\left. +\left( \Delta -1\right) ^{\frac{\nu -2}{2}}+\frac{(2+\nu )\left(
\Delta -1\right) ^{\frac{\nu -2}{2}}}{\Delta }\left( 2-\Delta \right) \right]
,  \notag \\
\Delta ^{\prime } &=&-4\alpha \left[ 1+\left( 1-k(\beta )^{2}\sin ^{2}\phi
\right) ^{\alpha }\right] \left( 1-k(\beta )^{2}\sin ^{2}\phi \right)
^{\alpha -1}kk^{\prime },  \notag \\
\frac{\Delta ^{\prime \prime }}{-4\alpha } &=&\left( 1-k(\beta )^{2}\sin
^{2}\phi \right) ^{\alpha -1}kk^{\prime }\left[ 2kk^{\prime }\left( \alpha
\left( 1-k(\beta )^{2}\sin ^{2}\phi \right) ^{\alpha -1}+\right. \right. 
\notag \\
&&\left. +\frac{\left[ 1+\left( 1-k(\beta )^{2}\sin ^{2}\phi \right)
^{\alpha }\right] (\alpha -1)}{1-k(\beta )^{2}\sin ^{2}\phi }\right) + 
\notag \\
&&\left. -\left[ 1+\left( 1-k(\beta )^{2}\sin ^{2}\phi \right) ^{\alpha }%
\right] \left( \frac{k^{\prime }}{k}+\frac{k^{\prime \prime }}{k^{\prime }}%
\right) \right] ,  \notag \\
N &=&1+\sqrt{1-\left[ 2\frac{\sqrt{\Delta (\phi )-1}}{\Delta (\phi )}\right]
^{\nu }\sin ^{2}\varphi },\quad 1\leq N\leq 2  \notag
\end{eqnarray}%
where $\Delta $ and $c(\Delta )$ only depend on $\beta $ and $\phi $. It is
worth to note that the inner integrals on $\varphi $ can be computed in
terms of generalized elliptic functions: they are formally similar to the
term occurring in the two-dimensional model provided one substitute%
\begin{equation*}
k^{2}\rightarrow K_{\nu }.
\end{equation*}%
Thus, by treating $\sin ^{2}\varphi $, $I_{1}$ and $I_{2}$ as constants%
\footnote{%
Being $1\leq N\leq 2$ bounded, the $I_{i}$ are also bounded and therefore
they do not affect the critical behavior. Moreover, sin$^{2}\varphi $
appears in the numerator so that, to understand the structure of the
possible divergences, sin$^{2}\varphi $\ can be treated as a constant.}, one
can rewrite $C_{(3)}$\ as follows%
\begin{eqnarray}
C_{(3)} &\sim &\dint\limits_{0}^{\pi }d\phi \left\{ \left[ \Delta ^{\prime
}c\left( \Delta \right) \right] ^{2}\left[ I_{2}E^{(2)}\left( K_{\nu
}\right) -\frac{I_{1}}{2}E^{(3)}\left( K_{\nu }\right) \right] \right. + 
\notag \\
&&+\left. I_{1}\left[ \Delta ^{\prime \prime }c(\Delta )+\left( \Delta
^{\prime }\right) ^{2}\frac{\partial c(\Delta )}{\partial \Delta }\right]
E^{(1)}\left( K_{\nu }\right) \right\} ,  \label{calspec} \\
E^{(n)}\left( K_{\nu }\right) &=&\dint\limits_{0}^{\pi }d\varphi \frac{1}{%
\left( 1-K_{\nu }\sin ^{2}\varphi \right) ^{n/2}}.  \notag
\end{eqnarray}%
It is worth to note that $E^{(1)}$, $E^{(2)}$ and $E^{(3)}$ diverge when $%
K_{\nu }$\ approaches to $1$. However, in the first two terms%
\begin{equation*}
K_{\nu }\rightarrow 1\Rightarrow c\left( \Delta \right) \rightarrow 0
\end{equation*}%
so that the divergences of the elliptic functions is canceled by the zeros
of $c\left( \Delta \right) $. One can see this as follows%
\begin{equation*}
E^{(n)}\left( K_{\nu }\right) \underset{K_{\nu }\rightarrow 2}{\sim }\frac{1%
}{\left\vert K_{\nu }-1\right\vert ^{a_{n}}},\quad n=2,3,\quad
a_{1}<a_{2}=1\Rightarrow
\end{equation*}%
\begin{eqnarray*}
\frac{1}{\left\vert K_{\nu }-1\right\vert ^{a_{n}}} &\sim &\left\vert \frac{%
\left( \sqrt{1+x}\right) ^{\nu }-\left( 1+\frac{x}{2}\right) ^{\nu }}{\left(
1+\frac{x}{2}\right) ^{\nu }}\right\vert ^{-a_{n}}\underset{x\rightarrow 0}{%
\sim }\frac{1}{x^{b_{n}}} \\
x &=&\left( \Delta -2\right) ,\quad b_{n}\geq 0.
\end{eqnarray*}%
$\nu $ has to be such that the product%
\begin{equation*}
\left\vert \left( c(\Delta )\right) ^{2}E^{(n)}\left( K_{\nu }\right)
\right\vert \sim \left\vert \frac{x^{2}}{\left\vert \left( \sqrt{1+x}\right)
^{\nu }-\left( 1+\frac{x}{2}\right) ^{\nu }\right\vert }\right\vert <\infty
\end{equation*}%
is finite for $\left( \Delta -2\right) $ approaching to zero so that $\nu $
has to fulfil%
\begin{equation*}
\nu \geq \nu ^{\ast }\geq 1.
\end{equation*}%
This constraint, as it will be pointed out in the next section, can be
interpreted as a positive self-consistency check of the present scheme. In
the last term in Eq. (\ref{calspec}) there is a logaritmic divergence when $%
K_{\nu }$\ approaches to $1$%
\begin{equation*}
E^{(1)}\left( K_{\nu }\right) \underset{\Delta \rightarrow 2}{\sim }\log
\left\vert \Delta -2\right\vert
\end{equation*}%
and this is not compensated by $c\left( \Delta \right) $. In fact, the outer
integral in $\phi $ smooths out this divergence: in the potentially
dangerous region such a contribution is 
\begin{equation*}
\dint\limits_{0}^{\pi }d\phi E^{(1)}\left( K_{\nu }\right) \underset{\Delta
\rightarrow 2}{\sim }\dint\limits_{0}^{\pi }d\phi \log \left\vert \Delta
-2\right\vert \underset{\Delta \rightarrow 2}{\sim }
\end{equation*}%
\begin{equation*}
\underset{\Delta \rightarrow 2}{\sim }\dint\limits_{\phi ^{\ast
}-\varepsilon }^{\phi ^{\ast }+\varepsilon }d\phi \log \left\vert \left(
1-a_{1}\sin ^{2}\phi \right) ^{b_{2}}-a_{2}\right\vert <\infty ,
\end{equation*}%
where $\phi ^{\ast }$ is the angle such that the argument of the logaritm
vanishes and $a_{1}$ and $a_{2}$ are two real constants. It is then apparent
that the divergences comes from the terms $\Delta ^{\prime \prime }$ and $%
\Delta ^{\prime }$: such divergences indeed occur for $k$ approaching to $1$%
. The term which dominates in this limit is the one coming from $\Delta
^{\prime \prime }$ (such a term of $\Delta ^{\prime \prime }$ is the one in
which there is no factor $k^{\prime }$) its contribution being%
\begin{equation*}
C_{(3)}\sim \dint\limits_{0}^{\pi }d\phi \frac{1}{\left( 1-k^{2}\sin
^{2}\phi \right) ^{1-\alpha }}.
\end{equation*}%
Therefore, one can choose the parameter $\alpha $ in order to obtain the
correct critical exponent. It is interesting that the qualitative behaviors
for very high and very low temperatures are in qualitative agreement with
what one expects since the novel contribution to the free energy is bounded
away from the critical region in such a way that the qualitative behavior is
not very different from the 2D Ising model. The others free parameters (that
is, $\lambda $, $\nu $ and the overall scale of the temperature which has
not been explicitly written) do not influence the critical region. The
overall scale of the temperature should be fixed in such a way that $k$ is
equal to one at the critical temperature (which, of course, depends on the
dimension). In 3D it is known (see \cite{PV02}) but in any case the present
scheme (provided the parameters are suitably chosen) seems to be able to
describe the model in higher dimensions also (in case, with further
dressings).

\section{Exploring high temperatures}

It is an interesting results that a very natural ansatze like the one
proposed in the present approach is able to describe the critical region of
the 3D Ising model with a single free parameter. This result supports the
use of Regge theory (and, more generally, of the "old-fashioned" tools of
strong interactions based on the $S$ matrix and its analytical properties;
see, for two detailed review, \cite{Co71}, \cite{Ve74}) in statistical
mechanics. The scheme proposed here has a methodological interest in itself
and could be useful in other situations. For instance, the \textit{%
Sherrington-Kirkpatrick} solution of the mean field spin glass model (as
well as, in many respects, the full Parisi solutions) can be naturally
interpreted as "dressing" of the trivial mean field spin glass model with a
highly non trivial expression for the dressing functions giving the spin
glass quantities when acting on the mean field quantities (see, for
instance, \cite{MPV87}, \cite{Ni02}\ and \cite{CC05}). There are many
quantities of interest in the mean field spin glass model (such as the
magnetization, the free energy and so on) \ which can be written
schematically as follows%
\begin{equation*}
m_{SG}(\beta )=\int Dz\rho _{SG}(z,\beta )m_{mf}(z)
\end{equation*}%
where $m_{SG}(\beta )$ is the magnetization of the spin glass mean field
model, $m_{mf}(z)$ is the magnetization of the standard mean field model
without disorder and $\rho _{SG}(z,\beta )$\ is a suitable highly non trivial%
\footnote{%
To get an idea of the difficulty of the problem, one can think that the
discovery of Parisi has been done in \cite{Pa79}. Only recently \cite%
{Gu02,GT02,Ta03}, 23 years later, it has been possible to prove rigorously
that the Parisi solution is the right one to describe the mean field glassy
phase.} dressing function whose explicit expression is known (see, for
instance, \cite{Ni02}). The above dressing function encodes non trivial
information about the physical effects of "glassy" disorder. The present
method suggests a new way to write down suitable ansatze to describe non
mean field spin glass systems. One could dress the free energy or the
magnetization of the 2D Ising model with function(al)s inspired by the
corresponding dressing function $\rho _{SG}(z,\beta )$\ of the mean field
model in which some parameters should be introduced in order to take into
account that one does not expect that exactly the same function describes
both the mean field and the non mean field case. In the spirit of the Regge
approach to particles physics, the few parameters which one introduces could
be fitted on experimental and/or numerical data to provide with a detailed
description of the much harder non mean field spin glass model. To be more
concrete, without entering into the subtle details of the theory of mean
field spin glass theory (which can be found, for instance, in \cite{MPV87}, 
\cite{Ni02}\ and \cite{CC05}), the free energy $F_{SG}$ of the mean field 
\textit{Sherrington-Kirkpatrick} model is%
\begin{equation}
F_{SG}(\beta ,h)=\underset{x}{\inf }(\log 2+f(0,h;x,\beta )-\frac{\beta ^{2}%
}{2}\int_{0}^{1}qx(q)dq)  \label{spgl1}
\end{equation}%
where $h$ is the magnetic field, $x$ is the spin glass functional order
parameter (which is a non decreasing function of an auxiliary variable $q$
such that both $x$ and $q$ take value into $\left[ 0,1\right] $). The
function $f(q,y;x,\beta )$ has a very important role and its dependence on
the variables $q$ and $y$ is determined by the equation%
\begin{equation}
\partial _{q}f+\frac{1}{2}\partial _{y}^{2}f+\frac{x(q)}{2}\left( \partial
_{y}f\right) ^{2}=0  \label{spgl2}
\end{equation}%
with final condition%
\begin{equation*}
f(1,y)=\log \cosh (\beta y).
\end{equation*}%
The dependence of $f$ on $\beta $ and $x$ is not important for the present
discussion. In any case, the spin glass formalism gives rise to a non
trivial "dressing" of the trivial thermodynamic function $\log \cosh (\beta
y)$ through the above non linear anti-parabolic equation. Therefore, as a
first attempt to attach the problem of non mean field spin glass systems
(and, in particular, the 2 dimensional Ising spin glass) one could change
the final condition of Eq. (\ref{spgl2}) and using as final condition,
instead of the trivial thermodynamic function $\log \cosh (\beta y)$, a
function similar to the free energy of the 2 dimensional Ising model such as%
\begin{equation*}
f_{\alpha ,\lambda }(1,y)=\log \cosh (\beta y)+\frac{\lambda }{2\pi }%
\dint\limits_{0}^{\pi }dx\log \left\{ \frac{1}{2}\left[ 1+\left( 1-k(\beta
y)^{2}\sin ^{2}x\right) ^{\alpha }\right] \right\}
\end{equation*}%
(in which two parameters $\alpha $ and $\lambda $ have been introduced in
order to allow the possibility to use both $\alpha $ and $\lambda $ as
variational parameters). Indeed, more general ansatze are also possible, the
basic physical idea remaining the same: according to the interpretation of
the glassy disorder as a "dressing" of a non glassy system, one can change
the initial data of Eq. (\ref{spgl2}) in such a way to obtain an effective
description of non mean field glassy systems. I hope to return on this
problem in a future publication.

The skeptical reader could simply tells that it is easy to describe the
critical behavior of a system by writing down a function with some free
parameters. On the other hand, it is also interesting to notice that the
present scheme could also be suitable to describe the model in a wider range
with the same few parameters. Let us consider for instance the small $\beta $
region. The first coefficients of the Taylor expansion of $F_{3D}^{(\nu
,\lambda =1)}$ for small $\beta $ read%
\begin{equation*}
\left( \frac{4\pi ^{2}F_{3D}^{(\nu ,\lambda =1)}}{\lambda }\right) _{\beta
\beta }=\int \int d\phi d\varphi \left[ \left( \frac{-1}{N^{2}}\right)
\left( I_{(1)}\right) ^{2}+\frac{I_{(2)}}{2N}+\frac{I_{(3)}}{2N}\right]
\end{equation*}%
\begin{eqnarray*}
I_{(1)} &=&\left( -\frac{\partial }{\partial \Delta }f^{(\nu )}\right) \frac{%
\Delta ^{\prime }}{\sqrt{1-f^{(\nu )}}},\quad I_{(2)}=\left( -\frac{\partial 
}{\partial \Delta }f^{(\nu )}\right) \frac{\Delta ^{\prime \prime }}{\sqrt{%
1-f^{(\nu )}}}, \\
I_{(3)} &=&\left( -\frac{\partial ^{2}}{\partial \Delta ^{2}}f^{(\nu
)}\right) \frac{\left( \Delta ^{\prime }\right) ^{2}}{\sqrt{1-f^{(\nu )}}}%
+\left( -\frac{\partial ^{2}}{\partial \Delta ^{2}}f^{(\nu )}\right) ^{2}%
\frac{\left( \Delta ^{\prime }\right) ^{2}}{2\left( \sqrt{1-f^{(\nu )}}%
\right) ^{3}}, \\
\Delta ^{\prime \prime } &=&\partial _{\beta }^{2}\Delta
\end{eqnarray*}%
\begin{equation*}
\left( \frac{4\pi ^{2}F_{3D}^{(\nu ,\lambda =1)}}{\lambda }\right) _{\beta
\beta \beta \beta }=\int \int d\phi d\varphi \left\{ \frac{-6}{N^{4}}\left(
I_{(1)}\right) ^{4}+\frac{2}{N^{3}}\partial _{\beta }\left[ \left(
I_{(1)}\right) ^{2}\right] +\frac{\left( I_{(1)}\right) ^{2}}{N^{3}}+\right.
\end{equation*}

\begin{eqnarray*}
&&-\frac{\partial _{\beta }\left( I_{(1)}I_{(2)}\right) }{2N^{2}}+\frac{%
\left( I_{(1)}\right) ^{2}\left( I_{(2)}+I_{(3)}\right) }{N^{3}}+\frac{1}{2N}%
\partial _{\beta }^{2}\left( I_{(2)}\right) + \\
&&\left. -\frac{I_{(1)}\partial _{\beta }\left( I_{(2)}\right) }{2N^{2}}+%
\frac{1}{2N}\partial _{\beta }^{2}\left( I_{(3)}\right) -\frac{%
I_{(1)}\partial _{\beta }\left( I_{(3)}\right) }{2N^{2}}-\frac{\partial
_{\beta }\left( I_{(1)}I_{(3)}\right) }{2N^{2}}\right\} .
\end{eqnarray*}%
The temperature enters in the above expressions only through the function $%
\Delta $ which is, of course, an even function of $\beta $ through the
function $k^{2}(\beta )$ which vanishes for $\beta =0$ together with its odd
derivatives. Therefore, in the expansion for small $\beta $ for the free
energy%
\begin{eqnarray}
\frac{4\pi ^{2}F_{3D}^{(\nu ,\lambda =1)}}{\lambda } &\sim
&\sum_{n=1}a_{2n}\beta ^{2n},  \notag \\
a_{2n} &=&\frac{1}{(2n)!}\frac{\partial ^{2n}}{\partial \beta ^{2n}}\left.
\left( \frac{4\pi ^{2}F_{3D}^{(\nu ,\lambda =1)}}{\lambda }\right)
\right\vert _{\beta =0}  \label{coefht}
\end{eqnarray}%
(in which the above coefficient have to be evaluated in $\beta =0$) many
terms do not contribute (such as the one in which at least one factor $%
I_{(1)}$ appear). In particular, the term which has the biggest numerical
coefficient is the one with the highest number of derivatives of $I_{(2)}$.
Generically, one gets%
\begin{equation*}
\partial _{\beta }^{2n-2}\left( I_{(2)}\right) \sim \partial _{\beta
}^{2n}\Delta .
\end{equation*}%
Such a factor is of order $(2n)!$ which compensates the denominators in Eq. (%
\ref{coefht}). There are further non vanishing terms in the derivatives of
the free energy: the number of these further terms increases with $n$.
However, the rate of increasing of the coefficients $a_{2n}$ should increase
very slowly with $n$. In other words, one should expect that the ratio%
\begin{equation}
\frac{a_{2n+2}}{a_{2n}}\sim tn^{c},\quad c>0,\quad t>0,\quad c\ll 1
\label{3Dnum}
\end{equation}%
where $c$ is a very small positive number. The reason is that all the terms
in which one takes derivative of $1/N$ are zero for $\beta =0$. All the
terms in which $I_{(1)}$ and/or $I_{(3)}$ appear without derivatives
vanishes for $\beta =0$ and so on. Consequently, the $a_{2n}$ cannot
increase fast with $n$. Let us consider the following function%
\begin{equation*}
\log (g(\beta ))
\end{equation*}%
where $g$ is an even function of $\beta $. It is possible to convince
oneself that if one compute 
\begin{equation*}
c_{2n}=\left. \partial _{\beta }^{2n}\log (g(\beta ))\right\vert _{\beta =0}
\end{equation*}%
only in one term it will be present the numerical factor $(2n)!$
compensating the denominator in Eq. (\ref{coefht}). Moreover, at order $2n$,
there are at most $n$ non vanishing terms contributing to $c_{2n}$ each of
which has a combinatorial factor\footnote{%
The subleading combinatorial factors are smaller of at least a factor $1/2n$
than the dominating combinatorial factor.} smaller than $(2n)!$. For these
reasons, the estimate (\ref{3Dnum}) could hold with $t$ of order $10$ and $c$
very close to zero. As far as the present scheme is concerned, the parameter
which is important in this regime is $\nu $ which is related to how fast the
free energy changes for small $\beta $.

It is worth to note that a too large $\nu $ (such as $\nu \gtrsim 10$) would
lead to a fast increase of the coefficient $a_{2n}$ in contrast with the
available numerical results (see, for instance, \cite{AF02}). As it has been
explained in the previous section, a value of $\nu $ lying, for instance, in
the interval 
\begin{equation*}
1\leq \nu \leq 2
\end{equation*}%
is compatible with the request to have the correct behavior in the critical
region. Thus, the constraints coming from the critical region are quite
consistent with the ones coming from the high temperatures region: something
which, \textit{a priori}, is far from being obvious.

These considerations seem to be in qualitative agreement with the observed
increase of the coefficients computed with numerical methods (see, for
instance, \cite{AF02}). In this scheme the constant $\lambda $ (which has no
role near the critical temperature) measure the "relative importance" of the
coefficients of the high temperature expansion of the 2D Ising model (which
is present in the complete expression Eq. (\ref{parto31})) and the genuine
3D term giving rise to the correct critical behavior. $\lambda $ also could
be extremely important in a numerical fit with the available data at high
temperature. Unfortunately, the present author is not expert in numerical
analysis. However, the above considerations suggest that it could be
possible to choose the parameters $\lambda $ and $\nu $ in order to achieve
a reasonable matching with the high temperature numerical results.

\section{Conclusion and perspective}

In this paper a Kallen-Lehman "phenomenological" approach to 3D Ising model
has been proposed. The approach is phenomenological in the sense that leads
to an analytic ansatze for the free energy in which there are free
parameters to be matched with experimental data and/or to be derived
theoretically (as it was the case in the Regge-S-matrix approach to strong
interactions). Such a scheme seems to be suitable to describe the critical
phase. Also the high temperature behavior could be captured by the present
scheme. It could be also interesting to apply the present scheme to non mean
field spin glass systems. Unfortunately, the present author is not expert in
numerical analysis. A careful numerical analysis, in case introducing few
more parameters, could shed light on the range of validity of the present
scheme. In any case, the present proposal could also have a methodological
interest in that similar techniques can be applied in other situations.

\section*{Acknowledgements}

The author would like to thank Prof. G. Vilasi and L. Parisi for continuous
encouraging and discussions and Prof. C. Martinez for important suggestions
and for convincing me in pursuing this idea. This work has been partially
supported by PRIN SINTESI 2004 and by Proy. FONDECYT N%
${{}^\circ}$%
3070055. 

\end{document}